\documentclass[a4paper]{PoS}

\pdfoutput=1

\usepackage{graphicx}
\usepackage{lineno}
\usepackage{mathtools}
\usepackage{subcaption}
\usepackage{units}

\newcommand*\patchAmsMathEnvironmentForLineno[1]{%
  \expandafter\let\csname old#1\expandafter\endcsname\csname #1\endcsname
  \expandafter\let\csname oldend#1\expandafter\endcsname\csname end#1\endcsname
  \renewenvironment{#1}%
     {\linenomath\csname old#1\endcsname}%
     {\csname oldend#1\endcsname\endlinenomath}}%
\newcommand*\patchBothAmsMathEnvironmentsForLineno[1]{%
  \patchAmsMathEnvironmentForLineno{#1}%
  \patchAmsMathEnvironmentForLineno{#1*}}%
\AtBeginDocument{%
\patchBothAmsMathEnvironmentsForLineno{equation}%
\patchBothAmsMathEnvironmentsForLineno{align}%
\patchBothAmsMathEnvironmentsForLineno{flalign}%
\patchBothAmsMathEnvironmentsForLineno{alignat}%
\patchBothAmsMathEnvironmentsForLineno{gather}%
\patchBothAmsMathEnvironmentsForLineno{multline}%
}

\newcommand{\annuluswidth}{\unit[10]{m}}

\newcommand{\fhit}{f_\text{hit}}

\newcommand{\correlationsubfig}[3]{
    \begin{subfigure}{0.5\textwidth}
        \includegraphics[width=\textwidth]{#1Correlation.pdf}
        \caption{#2}
        \label{fig:#3 correlation}
    \end{subfigure}
}

\newcommand{\collabel}[1]{\multicolumn{1}{c|}{#1}}

\title{Measuring High-Energy Spectra with HAWC}
\author{
    S. S. Marinelli$^a$ and \speaker{J. A. Goodman}$^b$ for the HAWC
    Collaboration\footnote{
        A complete list of authors is available at
        http://www.hawc-observatory.org/collaboration/icrc2017.php.
    }\\
    \llap{$^a$}Michigan State University, United States\\
    Email: \email{marine20@msu.edu}\\
    \llap{$^b$}University of Maryland, United States\\
    Email: \email{goodman@umdgrb.umd.edu}
}

\abstract{
    The High-Altitude Water-Cherenkov (HAWC) experiment is a TeV $\gamma$-ray
    observatory located \unit[4100]{m} above sea level on the Sierra Negra
    mountain in
    Puebla, Mexico. The detector consists of 300 water-filled tanks, each
    instrumented with 4 photomultiplier tubes that utilize the water-Cherenkov
    technique to detect atmospheric air showers produced by cosmic $\gamma$
    rays.
    Construction of HAWC was completed in March of 2015. The experiment's wide
    instantaneous field of view (\unit[2]{sr}) and high duty cycle (> 95\%)
    make it a
    powerful survey instrument sensitive to pulsars, supernova remnants, and
    other $\gamma$-ray sources. The mechanisms of particle acceleration at
    these
    sources can be studied by analyzing their high-energy spectra. To this end,
    we have developed an event-by-event energy-reconstruction algorithm using
    an artificial neural network to estimate energies of primary $\gamma$ rays
    at
    HAWC. We will present the details of this technique and its performance as
    well as the current progress toward using it to measure energy spectra of
    $\gamma$-ray sources.
}

\FullConference{
    35th International Cosmic Ray Conference---ICRC2017\\
    10--20 July, 2017\\
    Bexco, Busan, Korea
}
\ShortTitle{HAWC Spectral Measurements}

\begin{document}

    \section{Introduction}
    
        Among the fundamental questions in particle astrophysics are those of
        the
        sources and acceleration mechanisms of the high-energy particles that
        are observed to arrive at the Earth from space. Theories postulating
        different acceleration mechanisms have implications for the
        distributions of energies of particles emitted at these sources and
        therefore can be constrained by measurements of these sources' emitted
        energy spectra. Making such measurements requires the precise
        reconstruction of particle energies by observatories sensitive to these
        particles.
        
        The High-Altitude Water-Cherenkov (HAWC) detector observes TeV $\gamma$
        rays from an altitude of \unit[4100]{m} in the state of Puebla, Mexico.
        The experiment's 300 water-filled tanks are each instrumented with 4
        photomultiplier tubes (PMTs) that detect the Cherenkov light produced
        by
        $\gamma$-ray-induced extensive air showers as they pass through the
        tanks.
        Particle arrival times and deposited charges are reconstructed from PMT
        data,
        allowing estimation of various shower parameters such as zenith and
        azimuth angles and primary energy. A detailed
        description of the detector's electronics and data-acquisition system
        can be found at \cite{0004-637X-817-1-3}. By reconstructing primary
        energies with sufficient precision, HAWC can constrain the spectral
        energy distributions (SEDs) of $\gamma$-ray sources, thereby probing
        their
        mechanisms of particle acceleration. This paper introduces a new
        algorithm for reconstructing the energies of HAWC $\gamma$-ray events
        and
        establishes the performance of this technique via Monte Carlo (MC)
        simulations and on a calibration source, the Crab Nebula.
        
    \section{Energy-reconstruction technique}
    
        \label{sec:technique}
    
        The algorithm described herein uses an artificial neural network (NN)
        to
        reconstruct energies of photons detected by HAWC. It was implemented
        using the Toolkit for Multivariate Analysis \cite{TMVAShort}
        NN implementation.
        
        An NN is a complicated function mapping several quantities associated
        with an event (input variables) to some regression target or output
        variable, in this case $\log_{10} E$ where $E$ is the primary energy of
        the shower. A detailed description of the functional form of the NN can
        be found in \cite{TMVAShort}. This function is characterized by
        many (479 in this implementation) free parameters called weights. The
        optimal values of the weights are determined in a process called
        training, in which the weights are optimized using an MC
        simulation of the detector. The training process is described in
        Section \ref{sec:training}.
        
        The energy algorithm consists of two NNs: one for low-multiplicity
        events and one for high-multiplicity events. This binning has been
        found to yield better performance than a single NN even though
        multiplicity variables are included among the inputs.
        
        \subsection{Input variables}
        
            Fifteen input variables are used to characterize the shower energy.
            These variables have been chosen in order to capture three
            qualities of the shower: the particle multiplicity in the detector,
            the fraction of the shower which landed outside of the detector,
            and the atmospheric attenuation of the shower.
            
            The multiplicity in
            the detector is a crude measure of how much energy has been
            deposited in it and is quantified using the fraction of PMTs hit in
            the event, the fraction of tanks hit, and the normalization factor
            (in
            logarithm) of the fit of the lateral distribution of the shower.
            This fit is described in more detail at \cite{Abeysekara:2017mjj}.

            The distance between the reconstructed shower-core location and the
            center of the HAWC array is included as an input variable in order
            to provide information about how much of the shower is not
            contained within the array.
            
            Finally, the distance through which the
            shower has propagated in the atmosphere is quantified in two ways:
            using the reconstructed zenith angle and the lateral distribution
            of particles in the shower, which contains information about the
            shower's age. The latter is provided to the NN in the form of ten
            input variables, the $i$th of which is the fraction of
            charge deposition in the event
            occurring in PMTs whose distances from the shower axis, measured
            in the shower plane, are
            between
            $(\annuluswidth)i$ and $(\annuluswidth)(i + 1)$
            for $i$ between 0 and 8 inclusively; the last variable is the
            fraction of charge deposition in PMTs more than \unit[90]{m} from
            the
            axis.
            
        \subsection{Training}
        
            \label{sec:training}
            
            The goal of the training process is to choose the vector of NN
            weights $\mathbf{w}$ that minimizes the error function, defined as
            \begin{align}
                D\!\left(\mathbf{w}\right) \equiv \frac{1}{2}
                    \sum_{i=1}^{N_\text{events}} u_i \left[\log_{10}
                    \hat{E}\!\left(\mathbf{x}_i; \mathbf{w}\right) - \log_{10}
                    E_i\right]^2
            \end{align}
            where $u_i$ is the weight of the $i$th MC event (the
            event's relative
            importance in the training, unrelated to the NN parameters
            $\mathbf{w}$, which are
            unfortunately also called weights), $E_i$ is the true energy of the
            $i$th event, $\mathbf{x}_i$ is the $i$th event's vector of input
            variables, and $\hat{E}$ is the energy estimate for the event. This
            minimization is performed by repeatedly iterating over the sample
            of training events, with a small update to $\mathbf{w}$ performed
            for each event. The details of the
            Broyden-Fletcher-Goldfarb-Shanno algorithm for updating the
            weights can be found in \cite{TMVAShort}. The MC events are
            weighted (the $u_i$ are set) to resemble an $E^{-2}$ power-law
            spectrum during the training. This weighting was chosen because it
            produces
            a relatively
            flat RMS error between 1 and \unit[100]{TeV} (see Section
            \ref{sec:comparison}).
    
    \section{Performance on Monte Carlo}
    
        \label{sec:performance on MC}
    
        Several other energy-reconstruction algorithms have been developed
        for the HAWC observatory. These are described briefly in Section
        \ref{sec:other variables}, and the performance of all energy variables
        including the NN is described in Section
        \ref{sec:comparison}.
    
        \subsection{Other energy variables}
        
            \label{sec:other variables}
            
            The energy proxy that is presently used in the standard HAWC
            $\gamma$-ray analysis is the fraction of PMTs hit during the event,
            referred to as ``fraction hit'' or $\fhit$
            \cite{Abeysekara:2017mjj}.
            
            A
            likelihood-based algorithm for energy reconstruction has also been
            developed. This algorithm uses a likelihood function built using MC
            that takes into account the locations, arrival times, and deposited
            charges of the PMT hits that occur during the event. The hits are
            assumed to be statistically independent so that the likelihood is
            built by combining many single-hit likelihood functions. This
            algorithm has also been adapted for cosmic-ray energy
            reconstruction and is described in \cite{ZigCosmicRayICRC2017}.
            
            A third algorithm, called the ``ground parameter'' (GP), is an
            estimation
            of the
            photon primary energy as a function of reconstructed zenith angle
            and the value of the lateral distribution function at some
            optimal distance from the estimated shower axis. This is
            described in greater detail in \cite{KellyAPS2017}.
            
        \subsection{Performance comparison}
        
            \label{sec:comparison}
            
            The performance of these energy-reconstruction techniques is
            evaluated using MC simulations of $\gamma$-ray showers interacting
            with the HAWC detector. These simulations are described in
            \cite{Abeysekara:2017mjj}. The MC events are weighted to simulate a
            point source transiting at 20$^\circ$ declination with an
            $E^{-2.63}$ power-law energy spectrum. Analysis cuts are
            applied, including an $\fhit$ cut at 0.044 and a requirement that
            the
            reconstructed core be
            within
            the HAWC array. Background-rejection cuts are also applied in order
            to simulate a signal-event population resembling what would be used
            in an analysis of actual data.
        
            Figure \ref{fig:fraction-hit correlation} shows the distribution of
            fraction hit as a function of true energy.
            Because
            higher-energy photons on average produce higher-multiplicity
            showers,
            fraction hit is positively correlated with primary energy. However,
            several factors prevent it from determining the energy
            precisely. Because showers are attenuated as they travel through
            the atmosphere, a shower coming from a higher angle from zenith or
            having its first
            interaction higher in the atmosphere will have a lower multiplity
            than a shower of the same primary energy coming from a lower angle
            from zenith
            or interacting lower
            in the atmosphere. In addition, a shower whose core is not well
            centered on the detector will have a lower multiplicity in the data
            than one that is, again assuming equal primary energies.
            Furthermore, because fraction hit cannot exceed 1, at sufficiently
            high energies, the variable saturates and loses sensitivity to the
            energy.
            
            Figure \ref{fig:NN correlation} shows the relationship between the
            NN variable and true energy. Because the NN includes input
            variables that contain information about the containment of the
            shower within the array and its atmospheric attenuation, it is able
            to achieve a better resolution than that of fraction hit. Since
            the NN takes into account information beyond just the particle
            multiplicity of the shower, it does not saturate within
            the range of energies thrown in the MC, which extends well beyond
            \unit[100]{TeV}.

            \begin{figure}
                \correlationsubfig{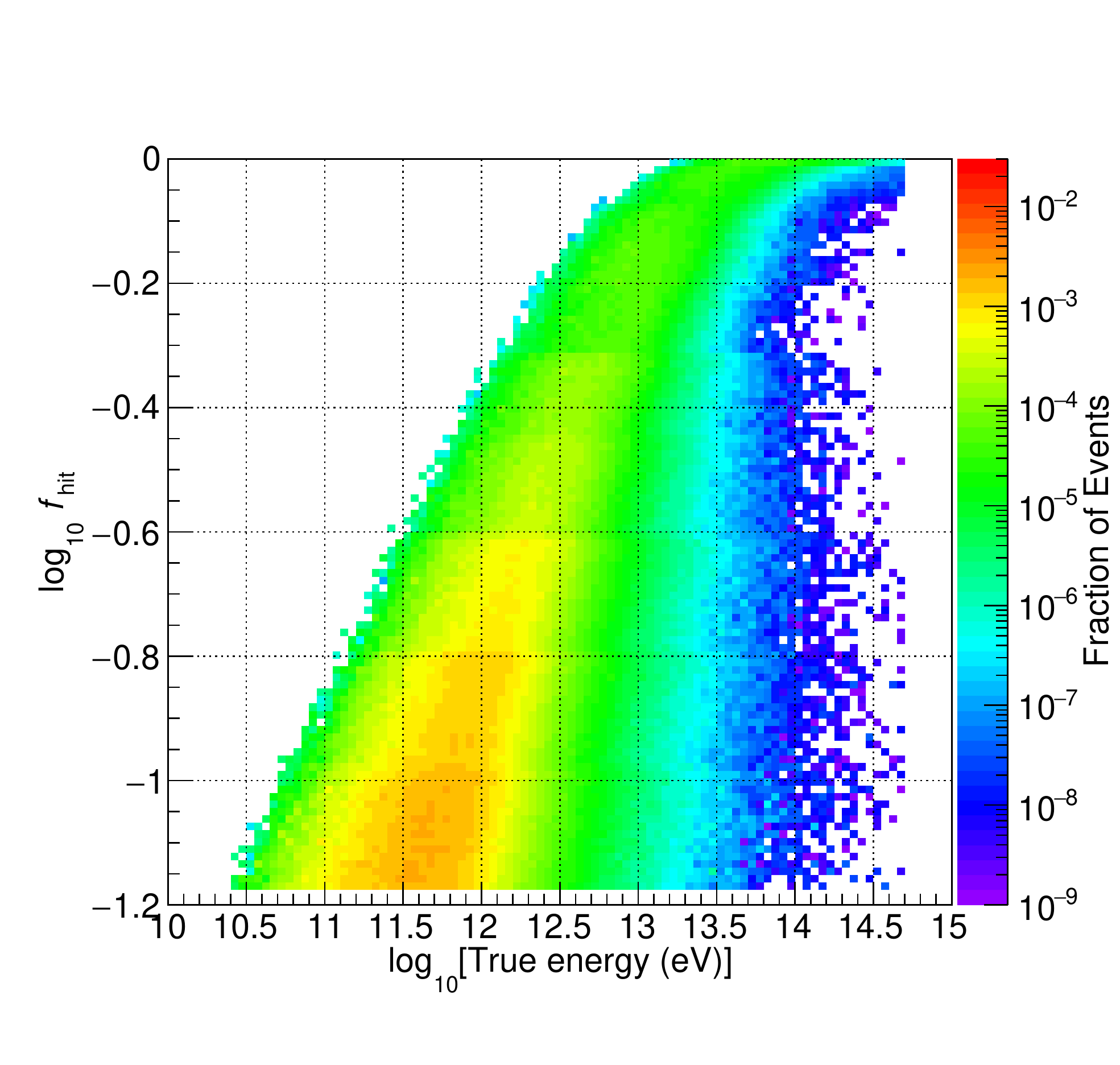}{Fraction hit}{fraction-hit}
                \correlationsubfig{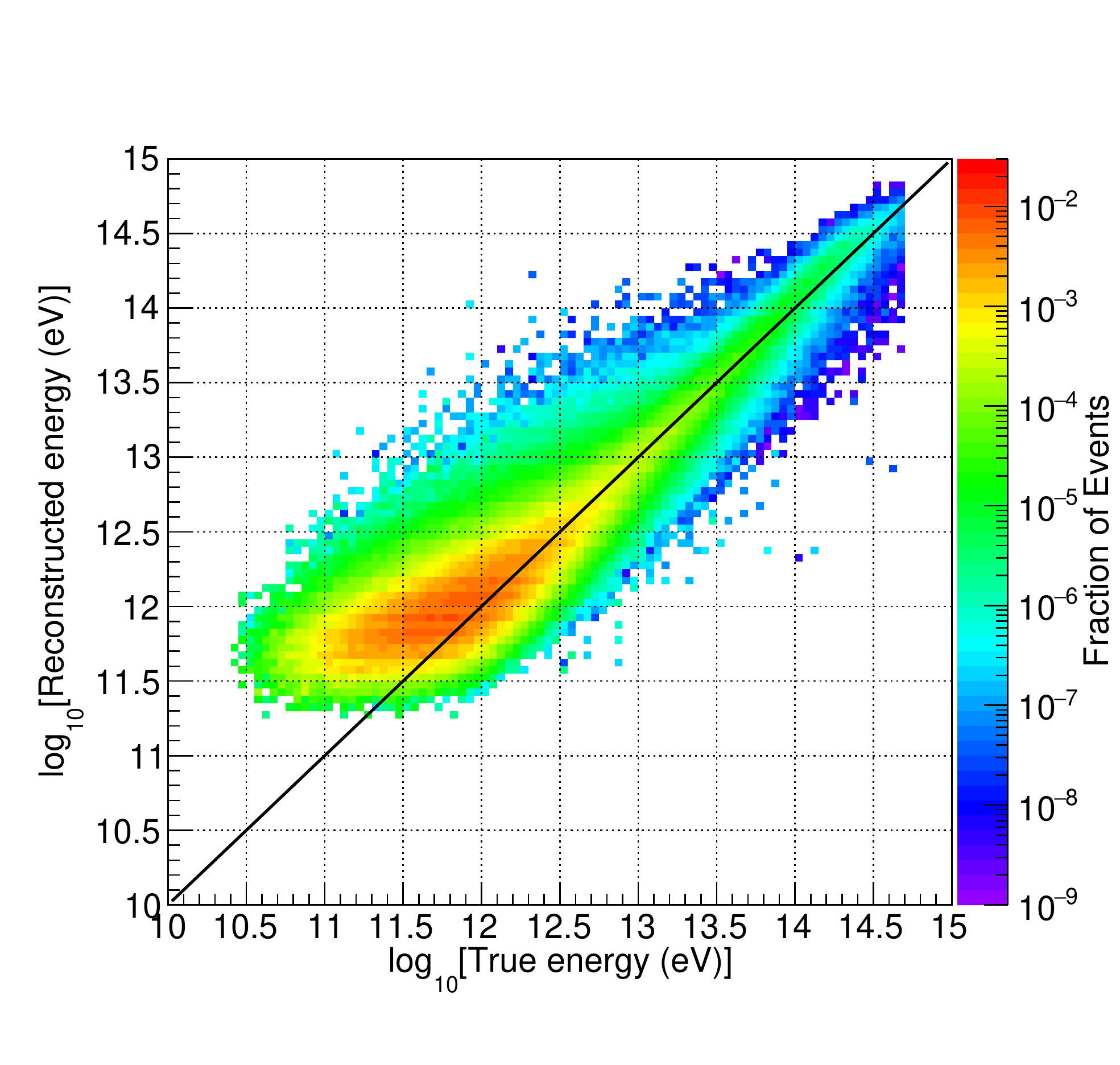}{NN estimator}{NN}
                \caption{
                    Joint distribution of each of two energy variables with
                    true
                    energy in
                    MC, among events passing the analysis cuts.
                }
            \end{figure}
            
\begin{figure}
                \includegraphics[width=\textwidth]{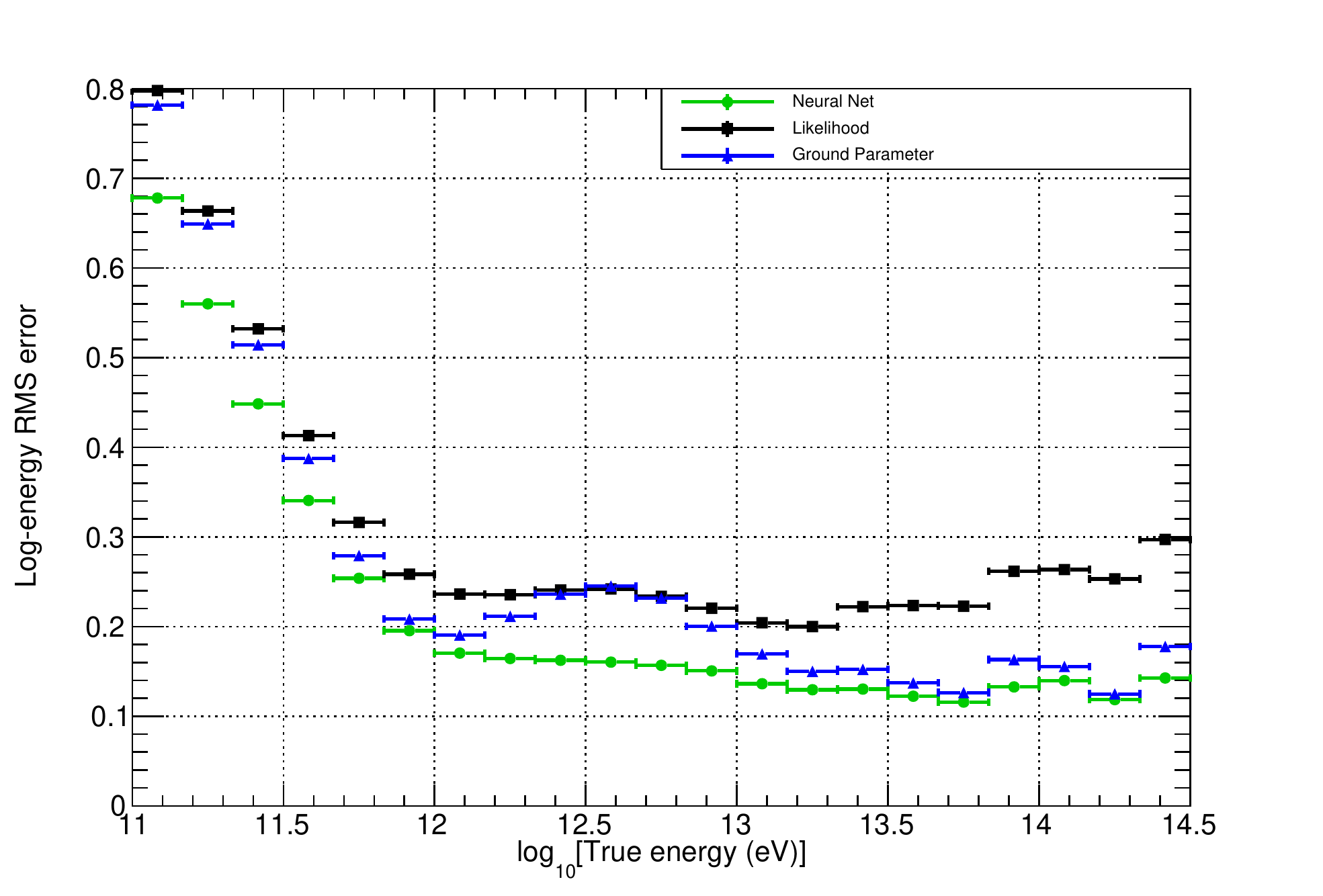}
                \caption{
                    RMS error in log space of the NN, likelihood, and
                    GP energy variables as a function
                    of true energy in MC.
                }
                \label{fig:RMS error}
            \end{figure}

            Figure \ref{fig:RMS error} shows the RMS error in log space of the
            three explicit energy estimates: the NN estimate, the likelihood
            estimate, and the GP estimate. The RMS error is defined as
            \begin{align}
                \rho \equiv \sqrt{\left\langle\left(\log_{10} \hat{E} -
                    \log_{10} E\right)^2\right\rangle}.
            \end{align}
            The NN displays the best RMS error in all bins. The RMS
            error in the 46 to \unit[68]{TeV} bin is around 0.12, indicating
            0.12 decade in energy, or about 32\% in linear energy
            space. All of the reconstruction algorithms perform poorly below
            around \unit[1]{TeV}. This is related to the $\fhit$ cut:
            a substantial fraction of events below \unit[1]{TeV} fail this cut,
            and those that pass it represent upwards multiplicity fluctuations.
            Thus their energies tend to be overestimated.
            
    \section{Calibration source: the Crab Nebula}
    
\begin{figure}
            \includegraphics[width=\textwidth]{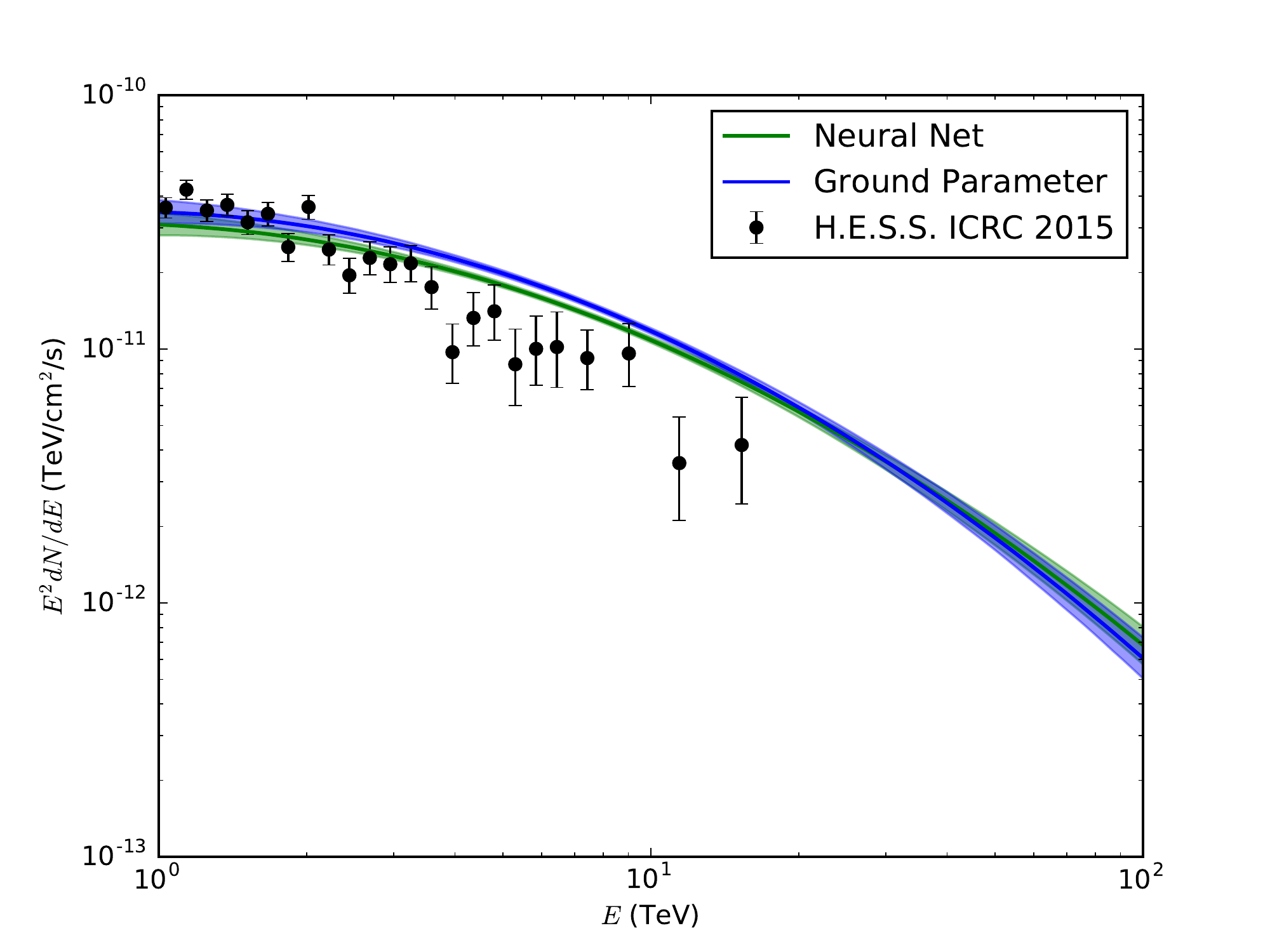}
            \caption{
                Preliminary fits of the Crab Nebula photon flux using the NN
                and
                GP energy variables. Shown errors on these fits are statistical
                only.
            }
            \label{fig:sed}
        \end{figure}

        In order to verify the ability of the new energy variables to
        accurately
        measure $\gamma$-ray energy spectra, a spectral fit of the Crab
        Nebula has
        been
        performed using the NN and GP variables. The Likelihood energy
        variable is not currently used in $\gamma$-ray analyses. This was
        carried
        out via the Multi-Mission
        Maximum Likelihood framework, described in \cite{Vianello:2015wwa},
        which
        interacts
        with HAWC's Analysis and Event-Reconstruction Integrated Environment
        to perform a forward-folded fit of the true-energy distribution to the
        observed energy-estimate distribution, taking into account the joint
        distribution in Figure \ref{fig:NN correlation}. The fit was performed
        using 18 months of HAWC data.
        The Crab was modeled as
        a point source, with the expected distribution of reconstructed photon
        directions being the detector's point-spread function centered on the
        source's location. Its energy
        spectrum was assumed to be log-parabolic,
        \begin{align}
            \frac{dN}{dE} = \Phi_0\left(\frac{E}{E_0}\right)^{-\alpha-\beta
                \ln\left(E/E_0\right)},
        \end{align}
        where $dN/dE$ is the photon particle flux, $\Phi_0$, $\alpha$,
        and $\beta$ are free parameters, and the pivot energy
        $E_0$ was chosen to be $\unit[7]{TeV}$ to decorrelate the estimates of
        $\Phi_0$ and $\alpha$. Events were sorted into two-dimensional bins of
        estimated energy and fraction hit in the analysis; fraction hit was
        included because it parameterizes the detector's angular resolution
        better than any of the explicit energy estimates. A skymap was
        constructed for each bin, and a pixel-by-pixel fit was performed to
        find
        the log-parabola spectrum most compatible with the data. The result is
        plotted in Figure \ref{fig:sed} along with H.E.S.S.'s reconstructed
        Crab spectrum from \cite{Holler:2015uca}, and the fit spectral
        parameters'
        values are given in Table \ref{tab:params} along with those from the
        published HAWC Crab spectrum in \cite{Abeysekara:2017mjj}.
    
        \begin{table}
            \centering
            \begin{tabular}{l|l|l|l|l|l|l|l}
                                     &                                          & \multicolumn{2}{c|}{Published HAWC}     & \multicolumn{2}{c|}{NN}                 & \multicolumn{2}{c}{GP}\\
                \collabel{Parameter} & \collabel{Unit}                          & \collabel{Value} & \collabel{Error}     & \collabel{Value} & \collabel{Error}     & \collabel{Value} & \multicolumn{1}{c}{Error}\\\hline
                $\Phi_0$             & \unit[$10^{-13}$]{(cm$^2$ s TeV)$^{-1}$} & 2.47             & $_{-0.12}^{+0.13}$   & 2.92             & 0.06                 & 3.22             & 0.07\\
                $\alpha$             &                                          & 2.627            & $_{-0.034}^{+0.035}$ & 2.712            & $_{-0.022}^{+0.021}$ & 2.751            & $_{-0.023}^{+0.022}$\\
                $\beta$              &                                          & 0.149            & $_{-0.030}^{+0.033}$ & 0.162            & 0.015                & 0.179            & $_{-0.016}^{+0.017}$
            \end{tabular}
            \caption{
                Best-fit values and statistical uncertainties for
                log-parabola spectral parameters.
            }
            \label{tab:params}
        \end{table}
        
        Table \ref{tab:params} shows that the NN spectral fit is
        better able to constrain all three spectral parameters than are
        the fit using only $\fhit$ and the GP fit.
        
        The SEDs computed using the two energy variables differ systematically
        from each other and from the H.E.S.S. result
        at some energies. An analysis of systematic errors has not yet been
        carried out and will be necessary in the future in order to understand
        this discrepancy. In particular the forward-folding technique is
        vulnerable to any differences in the distributions of the energy
        variables between data and MC, and such differences could result in
        the observed systematic discrepancy in the computed SEDs. The
        fractional systematic uncertainty on the published HAWC Crab
        spectral measurement
        is 50\% \cite{Abeysekara:2017mjj}, which, taken as a crude estimate of
        the systematic error on
        this analysis, is sufficient to account for the difference between the
        SEDs measured using the two new energy variables.
    
    \section{Conclusions}
    
        The development of the NN energy-reconstruction method allows HAWC
        to resolve energies up to \unit[100]{TeV}. The technique has already
        been put into use in measuring energy spectra of TeV blazars; see
        \cite{SaraAGNsICRC2017}.
        Using the NN method, HAWC
        will be able to measure energy spectra of gamma-ray sources up to
        unprecedentedly high
        energies, providing a new window into the high-energy universe.
        
    \acknowledgments
    
        We acknowledge support from the US National Science Foundation
        (NSF); the
        US Department of Energy Office of High-Energy Physics; the Laboratory
        Directed
        Research and Development (LDRD) program of Los Alamos National
        Laboratory;
        Consejo Nacional de Ciencia y Tecnolog\'{\i}a (CONACyT), M{\'e}xico
        (grants
        271051, 232656, 260378, 179588, 239762, 254964, 271737, 258865, 243290,
        132197), Laboratorio Nacional HAWC de rayos gamma; L'OREAL Fellowship
        for
        Women in Science 2014; Red HAWC, M{\'e}xico; DGAPA-UNAM (grants
        I100317,
        IN111315, IN111716-3, IA102715, 109916, IA102917); VIEP-BUAP; PIFI
        2012, 2013,
        PROFOCIE 2014, 2015; the University of Wisconsin Alumni Research
        Foundation;
        the Institute of Geophysics, Planetary Physics, and Signatures at Los
        Alamos
        National Laboratory; Polish Science Centre grant DEC-2014/13/B/ST9/945;
        Coordinaci{\'o}n de la Investigaci{\'o}n Cient\'{\i}fica de la
        Universidad
        Michoacana. Thanks to Luciano D\'{\i}az and Eduardo Murrieta for
        technical support.

    \bibliographystyle{JHEP}
    \bibliography{References}
    
\end{document}